# On transfer reactions playing a role in fission and in heavy-ion physics

G.Mouze, S.Hachem and C. Ythier

Faculté des Sciences, Université de Nice Sophia-Antipolis, France.

mouze@unice.fr

 **Abstract**

The cross section curves for the formation, at the barrier, of trans-target isotopes of a heavy element by bombardment of a heavy target with various heavy ions are shown to be similar to the distribution of the neutron number N  of a fission fragment around its most probable value $\bar{N}$. Moreover the isotopic cross sections for one-, two- and three- proton transfer products are found to be in agreement with the Gaussian distribution law of the atomic number Z of a fission product around its most probable value $\bar{Z}$. This situation suggests that the law of transfer of nucleons could be the same in fission and in particular heavy-ion reactions, and that the transfer time could be the same, i.e. of the order of 0.17 yoctosecond.

PACS numbers:

25.70.Hi: Transfer reactions

25.85.− w: Fission reactions

## 1. Introduction

In 2008 new ideas concerning the true nature of the reaction of nuclear fission were presented [1,2].

In particular the rearrangement step of the reaction is interpreted as the "transfer process" of a constant number of "nucleons" during an extremely short reaction time of 0.17 yoctosecond [3]. As a consequence of the energy-time uncertainty relation, uncertainties in mass number A, neutron number N and atomic number Z should be attached to any fission fragment.

The present communication is an answer to the question: Do other fields of nuclear physics present similar transfer reactions occurring within so short a reaction time ?

## 2. Main properties of the nucleon phase of the fission reaction.

At the end of the ignition step of the fission reaction in which a core-cluster system, e.g.

$$^{208}Pb + {}^{32}Mg + 79.36 \text{ MeV} \tag{1}$$

in the n-induced fission of $^{239}$Pu, has been formed, a core-cluster collision destroys the lead core, creates a hard *A = 126 nucleon-core* and releases <u>82</u> free nucleons [4]: a "nucleon phase" has been created, in which these <u>82</u> nucleons will be shared out, *by transfer*, between the cluster and the A = 126 nucleon-core.

But the study of the mass distributions of asymmetrically fissioning systems [4] revealed that a second hard core, an *A = 82 nucleon core*, is immediately formed, in this transfer, around the cluster, and that the remaining nucleons, 32 in this $^{239}$Pu example, are finally shared out between the two hard cores playing the role of "nascent" heavy and light fragments.

In fact, the whole rearrangement occurs within 0.17 yoctosecond, the mean value, for the best known fissioning systems, of the total transfer time, as proved by the uncertainties *Δ A, ΔN and ΔZ* in the mass-, neutron- and atomic-numbers of the light and heavy fragments: these uncertainties were revealed for the first time by the analysis of the data of the Saclay experiment on the "cold fission" of $^{235}$U induced by thermal neutrons [5,6 ]

The milieu in which the transfer occurs deserves to be called a nucleon phase, since *nucleon shells*, instead of the proton or neutron shells of ordinary nuclear matter, are closed there at magic 82 and 126 *mass* numbers, *as if any proton charge had momentarily disappeared.*

## 3. Isotopic yield distributions in fission and heavy-ion physics.

*3-1 The formation of heavy elements by transfer reaction*

In the eighties several groups of nuclear chemists attempted to create heavy nuclei by transfer of heavy ions, in the hope of creating new heavy elements. Such attempts were made in particular in Dubna [7], in Berkeley [ 8,9,10] and in Darmstadt [11].

For example, Schädel et al. report cross sections determined radiochemically for the heaviest known actinides, from Cf to Lr, produced in transfer reactions of $^{16,18}$O and $^{22}$Ne with $^{254}$Es as a target. Fig.1 shows , in particular, the isotopic distribution of the mendelevium isotopes formed by bombardment of $^{254}$Es with 127MeV$-^{22}$Ne ions (solid diamonds). One observes that the distribution is symmetrical (fig.1), and that the highest yield corresponds to a $^{256}$Md isotope formed by

127Mev $-^{22}$Ne + $^{254}$Es → $^{256}$Md + $^{20}$O        (2)

i.e. by capture of 2 protons and zero neutron. The Q-value is equal to  (- 17.44±0.06) MeV.

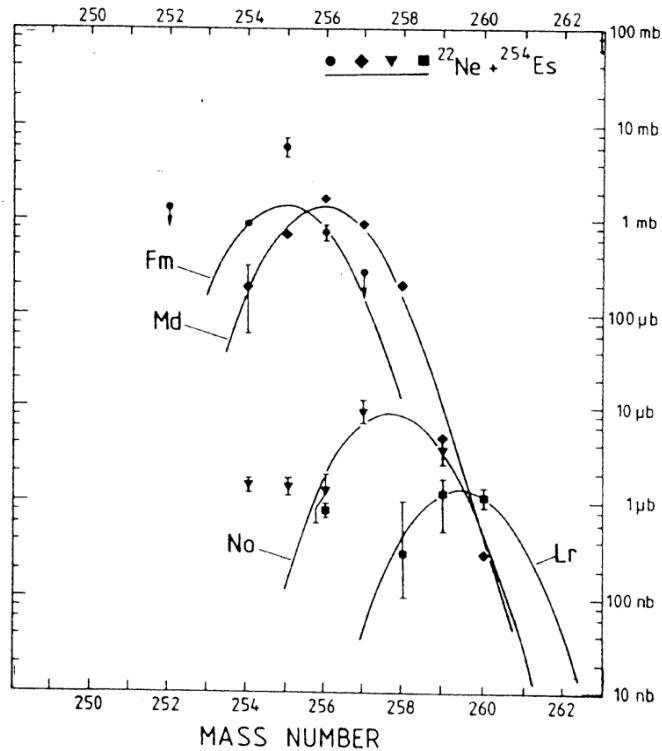

Fig.1 Isotopic distributions measured for 127 MeV $-^{22}$Ne on $^{254}$Es, according to Schädel et al. reproduced from ref.10 ( with permission). The data point for $^{260}$Md is from ref.[19].

Reaction (2) occurs, according to the authors of ref.10,at the Coulomb barrier, since the center- of- mass energy of the projectile of 108 MeV corresponds to its estimated value $B_c$ = 108 MeV; in such conditions, the excitation energy of the reaction products is very small, and competition with fission should have been avoided.

But the reported results raise many questions: Why are so many isotopes, and not only $^{256}$Md, formed in reaction (2)? and why are the yields distributed symmetrically around that of $^{256}$Md ? Schädel et al. propose a width of only 2.25 atomic mass units for the width of this distribution; however, the generally reported value for the width of isotopic distributions of trans-target elements formed in transfer reactions is 2.5 units [8,9,11]. But why have all these distributions the same width of 2.5 units? Should the Gaussian distribution be merely assigned to "damped collisions, in which the kinetic energies are relaxed " [12] ?

*3-2 Comparison of fission and heavy- ion collision.*

Today we propose to compare this width with that found for the isotopic distributions observed in a "cold fission" experiment realized at Saclay [13,14] and reported in ref.[5].

At a total kinetic energy of the fragments near to the maximum value, these authors have obtained the isotopic distributions of the elements indium ($Z = 49$), tin ($Z = 50$), antimony ($Z = 51$) and tellurium ($Z = 52$).

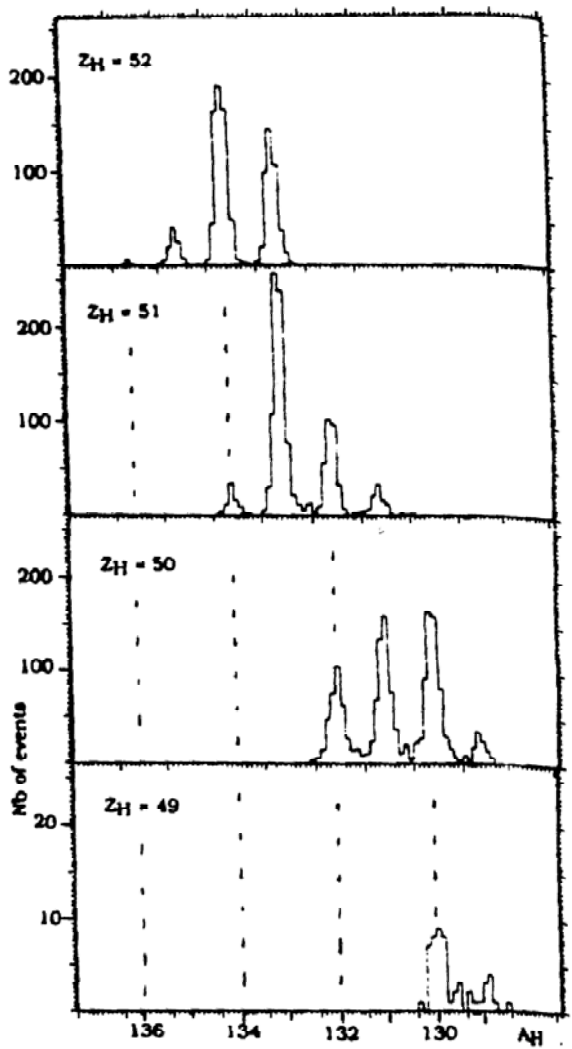

*Fig.2: Isotopic distributions of tellurium, antimony, tin and indium isotopes obtained in the n-induced "cold" fission of $^{235}U$, reproduced from ref.5 (with permission).*

The analysis of the histograms of the tin spectrum of fig.2, reported in ref.15, has shown that the yields are distributed on a Gaussian curve having a full width at half maximum (f.w.h.m.) of 2.54 u.

Moreover, this Gaussian curve has been interpreted, in this ref.15, as being nothing else but the distribution of the neutron numbers N of the tin isotopes about their most probable value $\overline{N}$ = 80.7.

*3-3 A common interpretation of both types of transfer reaction*

The similarity of the Gaussian curves obtained in fission and in transfer reactions is very striking. Thus, we are led to propose the following interpretation of fig.1: *this distribution is nothing else but the distribution of the neutron numbers N of the Mendelevium isotopes about their most probable value $\overline{N}$ = 155* (a value which might be a non-integer value).

In other words, we propose to consider that *a new state* of nuclear matter has been created in the collision of the $^{22}$Ne projectile with the $^{254}$Es nucleus. Its lifetime is of 0.17 yoctosecond. The resulting uncertainty ΔA in the mass A of the reaction product is of 4.175 mass units [3].

Since the transfer reaction occurs in a $^{254}$Es nucleus (Z = 99, N =155), the uncertainties in Z and N corresponding to the uncertainty in A, ΔA = 4.175 u, are ΔZ = ΔA (Z/A) i.e. 0.3898 ΔA ≅ 1.63u, and ΔN = ΔA (N/A), i.e. 0.6102 ΔA ≅ 2.55u.

This ΔN value of 2.55 u is in good agreement with all the data concerning the production of trans –target elements by transfer reaction: it is a consequence of the extreme brevity of the transfer time of 0.17 ys.

It remains to show that the ΔZ -value is in agreement with the cross-sections of 1-,2- and 3-proton transfer reactions.

## 4. On the yield of 1-,2- and 3-proton transfer reactions

Let us consider the reaction of 98 MeV$-^{18}$O ions with a $^{254}$Es nuclei, studied by Schädel et al.[10].

In fig.3, reproduced from ref.[10], the corresponding yield curves for the formation of Fm-, Md-, No- and Lr- isotopes are represented by solid curves.

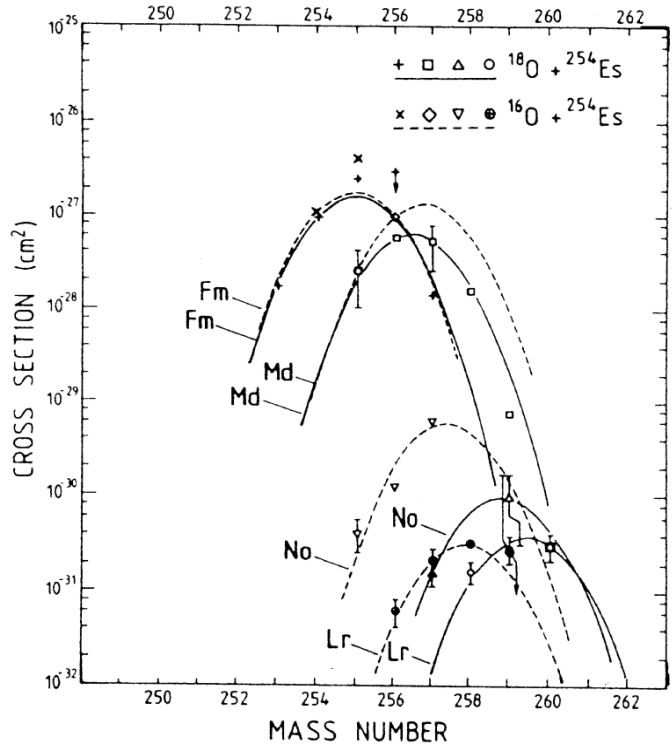

*Fig.3: Isotopic distributions measured for 101 MeV $-^{16}O$ and 98 MeV $-^{18}O$ on einsteinium 254, according to Schädel et al.,reproduced from ref.10 (with permission).*

But the relative value of the maximum yields of these elements can be predicted, if they are produced by transfer within 0.17 ys, on the basis of the uncertainty law in Z, i.e. on the basis of a Gaussian curve having a f.w.h.m. of 1.63 u, and representing *the distribution, as a function of the difference ($Z_i - Z_0$), of the probability P ($Z_i$) of creating the particular atomic number $Z_i$ in the transfer of i = 1, 2, 3… protons* to the $^{254}Es$ target ($Z_0 = 99$).

In particular the ratio of the cross sections for the formation of *fermium (Z= 100)*− and *nobelium (Z =102)− isotopes* can be estimated on the basis of this Gaussian curve having a f.w.h.m.  $\Delta Z = 1.63$ u.

Indeed, it can be shown that the expected yields are in the ratio ~ 4.238 $10^3$, since the ordinates of P($Z_i$), for  $Z_i = Z_0 + 1$, i.e. for $Z_1 = 100$, and for $Z_i = Z_0 + 3$, i.e. for $Z_3 = 102$, are in the ration (0.3934/9.2827 $10^{-5}$) [*] .

But according to the data furnished by Schädel et al.(fig.3), the maximum cross sections are respectively ~ 1.8 $10^{-27}$ cm$^2$ for fermium isotopes and ~ 0.95 $10^{-30}$ cm$^2$ for nobelium isotopes. They are also in the ratio ~ 1.895 $10^3$, i.e. almost in the expected ratio, within a factor of only 2.2.

---

[*] The ordinates $y_0/Y_1/y_2/y_3$ are in the ratio 1/0.3994/0.0170/9.2827 $10^{-5}$ for $\Delta Z = 1.63$.

This result shows that the *yields of the one- and three-proton transfer reactions are really determined by the law of uncertainty in the atomic number*.

Let us show that this law allows to determine the importance of the odd-even effect—an effect also reported by Scherrer et al. [11] in trans-target transfers—.

One observes indeed that the maximum cross sections for the formation of mendelevium and nobelium isotopes are not in the ratio

$R = (\sim 0.0170)/(\sim 9.2825 \ 10^{-5}) = \sim 1.83 \ 10^2$, predicted by the uncertainty law ,but in the ratio:

$R = (\sim 6.5 \ 10^{-28})/(\sim 0.95 \ 10^{-30}) = \sim 6.84 \ 10^2$.

This discrepancy results from the odd-even effect, i.e. from the greater probability of the formation of isotopes having an even Z.

One sees that the odd-even effect is of the order of $\sim 73$ % in two-proton transfer reactions.

## 5. Discussion

Isotopic distributions encountered in cold fission and in trans-target transfer reactions have a common characteristic both have a narrow width of 2.5 u. Let us show that this results from the fact that both correspond to "a selected single event".

Indeed, a selection has been realized in cold fission thanks to the choice of a narrow observation window "at the highest value of the total kinetic energy of the fission fragments".

Thanks to this selection, two discoveries were made possible: first, the mass yield curve of the *sole most energy-rich light fragment* of the n-induced fission of $^{235}$U, the $^{104}$Mo fragment, is a Gaussian curve having a width of 4.0 atomic mass units [6,15]; then, the mass yield curve of *the "element " tin* is a Gaussian curve having a f.w.h.m. of 2.5 u, as it result from fig.2.

In fact, without such a selection, one would, in fission, observe only mass yield curves belonging to *several* fragments, having also a width of at least 8 u, as it is the case for the mass distribution of $^{258}$Fm [4] , or  having a width of 30 u , as it is the case for the light- or heavy- fragment mass distribution, of the n-induced fission of $^{235}$U[4].

In transfer reactions, a similar selection has been obtained thanks to the choice of a "narrow observation window": the choice of a "one-, two- or three- proton transfer process".

Indeed, with such a selection, only one kind of transfer has occurred at the end of the nucleon phase, resulting in only one produced "most probable isotope" of a "single" element, e.g. the $^{256}$Md in the experiment of Schädel et al.[10]. Thus only one Gaussian distribution, having a *narrow* width of only 2.5 u at half maximum, could be observed, i.e. the distribution resulting from the uncertainty law in the neutron number at constant Z.

It is noteworthy that, without such a selection, isotopic distributions observed in transfer reactions would have a much wider width. Indeed, it has been reported that *isotopic distributions of below-target isotopes* have a considerable width at half maximum; e.g. Gäggeler et al. [16  ] report that in the 248-263 MeV $-^{48}$Ca + $^{248}$Cm reaction this width is of  5.0 u for the below Cm isotopes U and Pu, and is of 5.5 u for the below Cm isotopes from Rn to Th.

This last observation suggests that the wider width of below-target isotopic distributions could only result *from the addition of a small number of narrow single distributions* having the standard width of 2.5 u.

At this stage, we have shown why trans-target transfer reactions with heavy ions such as $^{16,18}$ O, $^{20,22}$ Ne, $^{40}$Ar, $^{40,44,48}$Ca, $^{136}$Xe or $^{238}$U all lead to quite similar Gaussian distributions of a  limited number of isotopes: Indeed, the nucleon phase explains the variation of their yield as a Gaussian curve corresponding to the uncertainty law in the neutron number.

Moreover, we have shown why the maximum yield of trans-target transfer of one, two and three protons decreases according to the uncertainty law in the atomic number: Indeed, the nucleon phase explains that the uncertainty in Z amounts to only  ~ 1.63 charge unit.

The absence of competition of fission can perhaps be explained. Let us recall that fission requires the destruction of the primordial hard $^{208}$Pb core of heavy nuclei. In the case of the $^{254}$Es target, clusterized into a $^{208}$Pb-$^{46}$Cl dinuclear system, the $^{46}$Cl cluster has a binding energy per nucleon of 8.102 keV, greater than the binding energy per nucleon of projectiles such as $^{18}$O (7,767 keV) or $^{22}$Ne (8,080 keV) [17], and it shares with $^{208}$Pb a considerable clusterization energy of 118 MeV. This situation explains why the projectile, rather than the $^{208}$Pb core, is destructed, at least partially, in the collision.

Authors have wondered that a shift of only two or three mass numbers are observed between $^{40}$Ca and $^{48}$Ca reactions, the eight neutron difference between the two kinds of projectiles being also partially reflected in the products [9]: Indeed, the sole nucleon phase decides on the "most probable " neutron number.

Authors have wondered that cross sections for the production of actinides from the transfer of the same nucleons [9], more precisely of the same number xp, yn of nucleons [10], are very similar for projectiles ranging from $^{18}$O to $^{238}$U: Indeed the study of the reaction of fission has shown that, in its rearrangement step, any distinction between proton and neutron has been abolished: *only "nucleons" are transferred*. There is no reason to believe that the situation is not the same in the formation of trans-target nuclei in heavy-ion transfer reactions. The "proton-character" reappears only at the end of the nucleon phase.

It would be interesting to find an explanation of the reported observation that the maximum yield of trans-target transfer reactions is obtained at the Coulomb barrier.

Last, but not least, it may be asked whether the above considerations can be extended to complete fusions of projectile and target nucleus.

Let us consider the reactions 243 MeV – $^{48}$Ca + $^{244}$Pu and 250 MeV – $^{48}$Ca + $^{244}$Pu. According to Oganessian et al.[20], they lead to the formation of $^{288}$(114) and $^{289}$(114) in the ratio 7/2 and 4/1, respectively. It means that the most probable values of the neutron number of the transfer product are ~ 174.3 and ~ 174.2, respectively, whereas the highest possible value of N would be 178, and correspond to the reaction

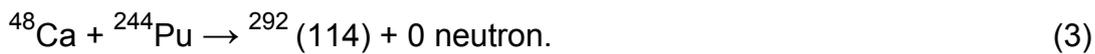

$$^{48}\text{Ca} + {}^{244}\text{Pu} \rightarrow {}^{292}(114) + 0 \text{ neutron.} \qquad (3)$$

At these projectile energies the number of emitted "prompt neutrons" should be $\bar{n} \cong$ 3.7 and $\bar{n} \cong 3.8$, respectively: they allow the formation of the most probable transfer product in the respect of the uncertainty law in N. The N-values are distributed on a Gaussian curve, having a f.w.h.m. of ~ 2.545 u if the transfer time is 0.17 ys: The competition of the two N-values proves the existence of this distribution.

# 6. Conclusion

There exists a great resemblance between isotopic distributions encountered in cold-fission experiments[†], on one hand, and those encountered in transfer reactions leading to trans-target nuclei, on the other hand. This resemblance suggests *that one and the same state of nuclear matter* intervenes in both cases. This state is referred to as "nucleon phase". Its lifetime is so short, only 0.17 yoctosecond, that uncertainties in A, in Z and in N are attached to the final transfer products.

We hope to have shown that these uncertainties explain the major properties of the distributions of the transfer products, either as a function of N, or as a function of Z.

---

[†] or in prompt gamma-ray coincidence experiments [15,18]

May these new points of view open the way to a better understanding of the state of nuclear matter under extreme conditions, and be useful for the synthesis of superheavy elements.